\documentclass[twocolumn,letterpaper]{IEEEAerospaceCLS}  

\usepackage{amsmath}
\usepackage{amssymb}
\usepackage{amsfonts}
\usepackage[]{graphicx}    
\newcommand{\ignore}[1]{}  
\usepackage{graphicx}
\usepackage{mfirstuc-english}
\usepackage{color}
\usepackage{bm}
\usepackage{hyperref}
\usepackage{subfig}
\hypersetup{hidelinks,pdfnewwindow,breaklinks} 
\MFUnocap{to}
\MFUnocap{on}
\MFUnocap{via}
\MFUnocap{under}
\MFUnocap{using}

\newcommand{\ie}{i.e.}

\newcommand{\myhref}[2]{\textcolor{uiucgrayblue}{\href{#1}{#2}}}
\definecolor{caltechgreen}{rgb}{0,0.3451,0.3137}
\definecolor{uiucgrayblue}{RGB}{111,175,199}

\let\oldmathbb\mathbb
\renewcommand{\mathbb}[1]{{\oldmathbb{#1}}}
\let\oldmathbf\mathbf
\renewcommand{\mathbf}[1]{{\oldmathbf{#1}}}
\let\oldmathcal\mathcal
\renewcommand{\mathcal}[1]{{\oldmathcal{#1}}}
\let\oldmathrm\mathrm
\renewcommand{\mathrm}[1]{{\oldmathrm{#1}}}

\usepackage{fancyhdr}

\usepackage{changepage}
\fancypagestyle{myfancy}{%
    \fancyhf{}  
    \fancyhead[LE]{\footnotesize {\textit{IEEE Aerospace Conference, Montana, 2023}}} 
    \fancyhead[RE]{\footnotesize \thepage} 
    \fancyhead[RO]{\footnotesize {\textit{Donitz et al.: Interstellar Object Accessibility and Mission Design}}} 
    \fancyhead[LO]{\footnotesize \thepage} 
}

\fancypagestyle{titlepagefancy}{
\fancyhf{}
\fancyhead[C]{\footnotesize \textbf{IEEE Aerospace Conference} \\
{Preprint Version, Accepted: November 2022 (DOI: {\color{uiucgrayblue}\href{https://doi.org/10.1109/AERO55745.2023.10115554}{https://doi.org/10.1109/AERO55745.2023.10115554}})\\
\vspace{3pt}
{\scriptsize © 2023 IEEE. All rights reserved. All requests for copying and permission to reprint should be submitted to {\color{uiucgrayblue}\href{www.copyright.com}{CCC}}.}}}
\fancyfoot[C]{\footnotesize \thepage} 
}

\begin{document}
\title{Interstellar Object Accessibility and Mission Design}

\author{%
\large{Benjamin P. S. Donitz$^*$~~~Declan Mages$^*$~~~Hiroyasu Tsukamoto$^\dag$~~~Peter Dixon$^*$} \vspace{0.3em}\\
\large{Damon Landau$^*$~~~Soon-Jo Chung$^{*\dag}$~~~Erica Bufanda$^\ddag$~~~Michel Ingham$^*$~~~Julie Castillo-Rogez$^*$}\vspace{1em}\\
\normalsize{$^*$Jet Propulsion Laboratory, California Institute of Technology, Pasadena, CA}\vspace{0.3em}\\
\normalsize{$^\dag$Division of Engineering and Applied Science, California Institute of Technology, Pasadena, CA}\vspace{0.3em}\\
\normalsize{$^\ddag$Institute for Astronomy, University of Hawaii at Manoa, Honolulu, HI}
\thanks{$^*$Jet Propulsion Laboratory, California Institute of Technology, Pasadena, CA, {\tt\footnotesize\{\href{mailto:benjamin.p.donitz@jpl.nasa.gov}{benjamin.p.donitz}, \href{mailto:declan.m.mages@jpl.nasa.gov}{declan.m.mages}, \href{mailto:peter.dixon@jpl.nasa.gov}{peter.dixon}, \href{mailto:damon.landau@jpl.nasa.gov}{damon.landau}, \href{mailto:michel.d.ingham@jpl.nasa.gov}{michel.d.ingham}, \href{mailto:julie.c.castillo@jpl.nasa.gov}{julie.c.castillo}\}@jpl.nasa.gov}. $^\dag$Division of Engineering and Applied Science, California Institute of Technology, Pasadena, CA, {\tt\footnotesize\{\href{mailto:htsukamoto@caltech.edu}{htsukamoto}, \href{mailto:sjchung@caltech.edu}{sjchung}\}@caltech.edu}. $^\ddag$University of Hawaii at Manoa, Honolulu, HI, {\tt\footnotesize\href{mailto:ebufanda@hawaii.edu}ebufanda@hawaii.edu}. Part of the research was carried out at the Jet Propulsion Laboratory, California Institute of Technology, under a contract with the National Aeronautics and Space Administration. Government sponsorship acknowledged.}
}

\maketitle

\thispagestyle{titlepagefancy}
\pagestyle{myfancy}

\begin{abstract}
Interstellar objects (ISOs) are fascinating and under-explored celestial objects, providing physical laboratories to understand the formation of our solar system and probe the composition and properties of material formed in exoplanetary systems. In this work, we investigate approaches to designing successful flyby missions to ISOs. We have generated trajectories to a series of synthetic representative ISOs, simulating a ground campaign to observe the target and resolve its state, and determining the cruise and close approach $\Delta V$ required for the encounter. We have developed novel deep learning-driven guidance and control algorithms to enable an accurate flyby of an ISO traveling at velocities over 60 km/s. In this paper, we discuss the accessibility of and mission design to ISOs with varying characteristics, including analysis of state covariance estimation throughout the cruise, handoffs from traditional navigation approaches to novel autonomous navigation for fast flyby regimes, and overall recommendations about preparing for the future in situ exploration of these targets.
\end{abstract}

\tableofcontents

\section{Introduction}
The identification of the first Interstellar Objects (ISOs) is one of the most impactful discoveries of the last decade. The science community, highlighted by the Planetary Science and Astrobiology Decadal Survey (PSADS), has continuously highlighted the importance of the exploration of ISOs \cite{NAP26522}. Though long theorized and expected to be abundant, so far, only three ISOs have been detected and confirmed: `Oumuamua in 2017 \cite{oumuamua}, Borisov in 2019 \cite{borisov}, and an interstellar meteor that was recently declassified in 2022 and impacted the Earth in 2014 \cite{ussf-memo}. 

The discovery of `Oumuamua and Borisov had a tantalizing effect on the planetary science community; observatories across the globe and in space trained their sensors on these fascinating targets to discern any available information on their origin, their composition, and their likeness to similar bodies that originated in our own solar system. Despite this concerted global observation effort, key questions about these ISOs remain open, and can only be addressed with a dedicated spacecraft flying by the ISO to obtain close-up/in-situ measurements of the ISO's shape, properties, and composition.

The discovery of `Oumuamua and Borisov triggered significant interest from the community in exploring these objects. The PSADS highlights ISOs as a strategic research objective, recommending that NASA "Identify and characterize interstellar objects…with spacecraft data, telescope observations, theoretical and modeling studies of their formation and evolution, and laboratory studies of analogue materials" \cite{NAP26522}. Next-generation observatories like the Vera C. Rubin observatory and NEO Surveyor are expected to dramatically increase the number of detected ISOs {hoover-iso}, increasing the prospect of exploring these bodies this decade if several challenges intrinsic to the exploration of these bodies can be resolved. This paper will focus on two of those challenges and discuss work performed as a collaboration between the NASA Jet Propulsion Laboratory and autonomous guidance and control (G\&C) research scientists at the California Institute of Technology to address these challenges. The first of these challenges is that ISOs are detected shortly prior to making a close approach to Earth, leaving little time for observatories on Earth to characterize and refine the ISO's trajectory prior to the launch of a spacecraft. The vehicle then must perform large maneuvers while on cruise to correct its trajectory as the ISO's position and orbit uncertainties decrease with longer observation arcs with Earth-based observatories. The second challenge is that ISOs, in general, travel with extremely high relative velocities, being that their hyperbolic velocity with respect to the Sun is greater than zero (not gravitationally bound). As the spacecraft approaches the ISO, the relative state changes dramatically, making it impossible to obtain a navigation image, communicate that image back to the ground for mission planning, and upload an updated set of commands for the spacecraft to perform the flyby accurately. The conditions are such that onboard autonomy is required for a successful flyby of the ISO.

To address both challenges, the authors developed and tested a novel approach to deep learning-driven autonomous navigation to fast flyby targets like ISOs. The authors tested that new G\&C algorithm on a population of 277 synthetic ISOs, each with a simulated campaign to determine the knowledge of the ISO's position over time while in cruise and during the autonomous navigation phase using optical navigation. The unification of ISO detection, orbit characterization, and cruise trajectory with learning-based G\&C algorithms for accurate low-$\Delta V$ flybys represents a nearly end-to-end simulation and assessment of a mission to visit an ISO. This process is simulated using JPL's SmallSat Development Testbed, which determines the feasibility of encountering ISOs with different initial detection and orbit characteristics. A paper previously presented at the 2022 IEEE conference discusses the method of determining the ISO position over time \cite{iso-2022} and a journal paper currently in pre-publication details the theory and application of the novel deep learning-based G\&C algorithm, Neural-Rendezvous~\cite{tsukamoto2022}. 

This paper will discuss the accessibility of and mission design to ISOs with varying characteristics, including a discussion of state covariance estimation throughout the cruise, handoffs from traditional navigation approaches to novel autonomous navigation for fast flyby regimes, and overall recommendations about preparing for future in-situ exploration of these targets. The lessons learned also apply to the fast flyby of other small bodies, including long-period comets and potentially hazardous asteroids, which also require tactical responses with similar characteristics.



\section{ISO Encounter Geometry}
\begin{figure}
    \centering
    \includegraphics[width=.45\textwidth]{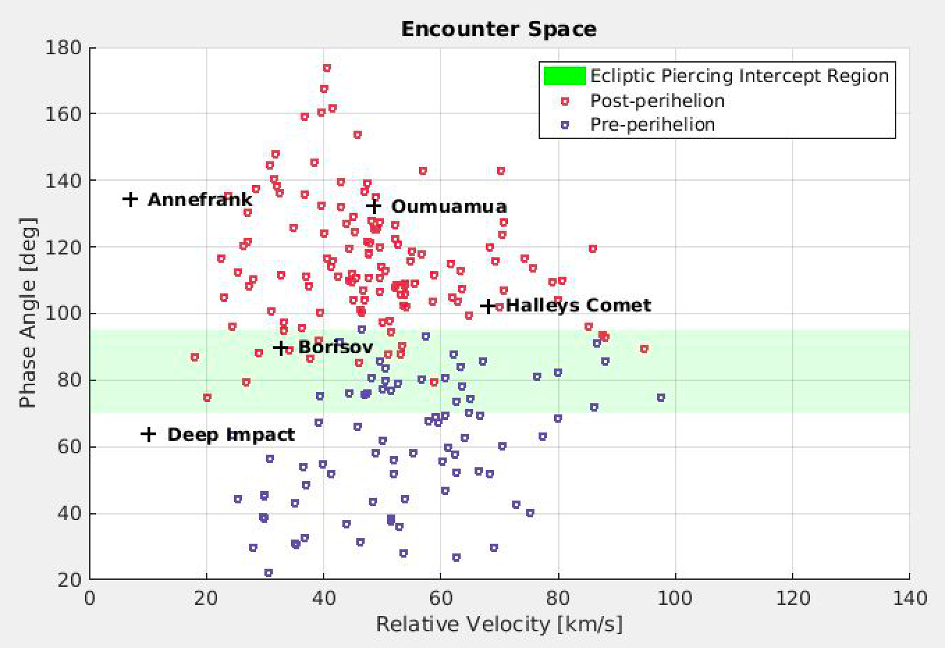}
    \caption{Most trajectories have relative velocities at $>$60 km/s and phase angles of $>$120$^\circ$.}
    \label{fig:vi_phase}
\end{figure}

Using a model developed by T. Engelhardt and J. Jedicke \cite{engelhardt}, the authors performed an extensive analysis of synthetic ISOs and their trajectories. The details of this analysis and its results are presented in \cite{iso-2022}. A key takeaway is that ISOs do tend to drive extremely fast flyby relative velocities. Figure \ref{fig:vi_phase} shows the distribution of all ISO flyby cases with their relative velocity and phase angle. A majority of the cases require relative flyby velocities over 60 km/s. For reference, many of the previous small bodies flybys have taken place at $<$30 km/s of this velocity (see Table \ref{tab:past_missions}). One notable exception to this is the European mission Giotto, which flew by Halley's Comet with a relative velocity of almost 70 km/s. 

\begin{table*}[]
\centering
\caption{Most previous small body flyby missions have occurred at relative velocities $<$30 km/s}
\label{tab:past_missions}
\begin{tabular}{|c|c|c|c|c|c|}
\hline
\textbf{Mission}    & \textbf{Target Body}  & \textbf{Relative Velocity (km/s)} & \textbf{Phase Angle (deg)} & \textbf{Flyby Distance (km)} & Source\\ \hline
\textbf{Deep Impact}  & Tempel 1 & 10.2                              & 64                         & 500      &  \cite{deep_impact_autonav}                  \\ \hline
\textbf{STARDUST}    & Wild 2 & 6.1                               & 72                         & 200         &    \cite{stardust}             \\ \hline
\textbf{Deep Space 1} & Borrelly & 16.6                              & 65                         & 2000     &  \cite{ds1}                  \\ \hline
\textbf{Giotto} & Halley's Comet & 68.4                              & 107                         & 596     &    \cite{giotto-1} \cite{giotto-2}               \\ \hline
\textbf{New Horizons} & Arrokoth & 14.3                              & 152                         & 3500    &      \cite{doi:10.1126/science.aaw9771}               \\ \hline
\end{tabular}%
\end{table*}

These high relative velocities introduce a variety of challenges, one of which is to navigation. For flybys of small bodies with poorly known ephemerides, optical navigation is a powerful tool to directly measure relative position using onboard sensors. Optical navigation provides angular measurements, and thus, the metric accuracy is directly proportional to the distance from the target; as the spacecraft gets closer, the knowledge of the ISO position improves. However with high relative flyby velocities, the relative distances are also high versus time, which means 24 hours from encounter, an optical navigation measurement may only be accurate to dozens of kilometers. After 24 hours it is increasingly difficult to have enough time to include ground in the loop, and so the only way to improve those accuracies is via autonomous optical navigation.

\section{Autonomous Navigation State of Practice}

\begin{figure*}
    \centering
    \includegraphics[width=\textwidth]{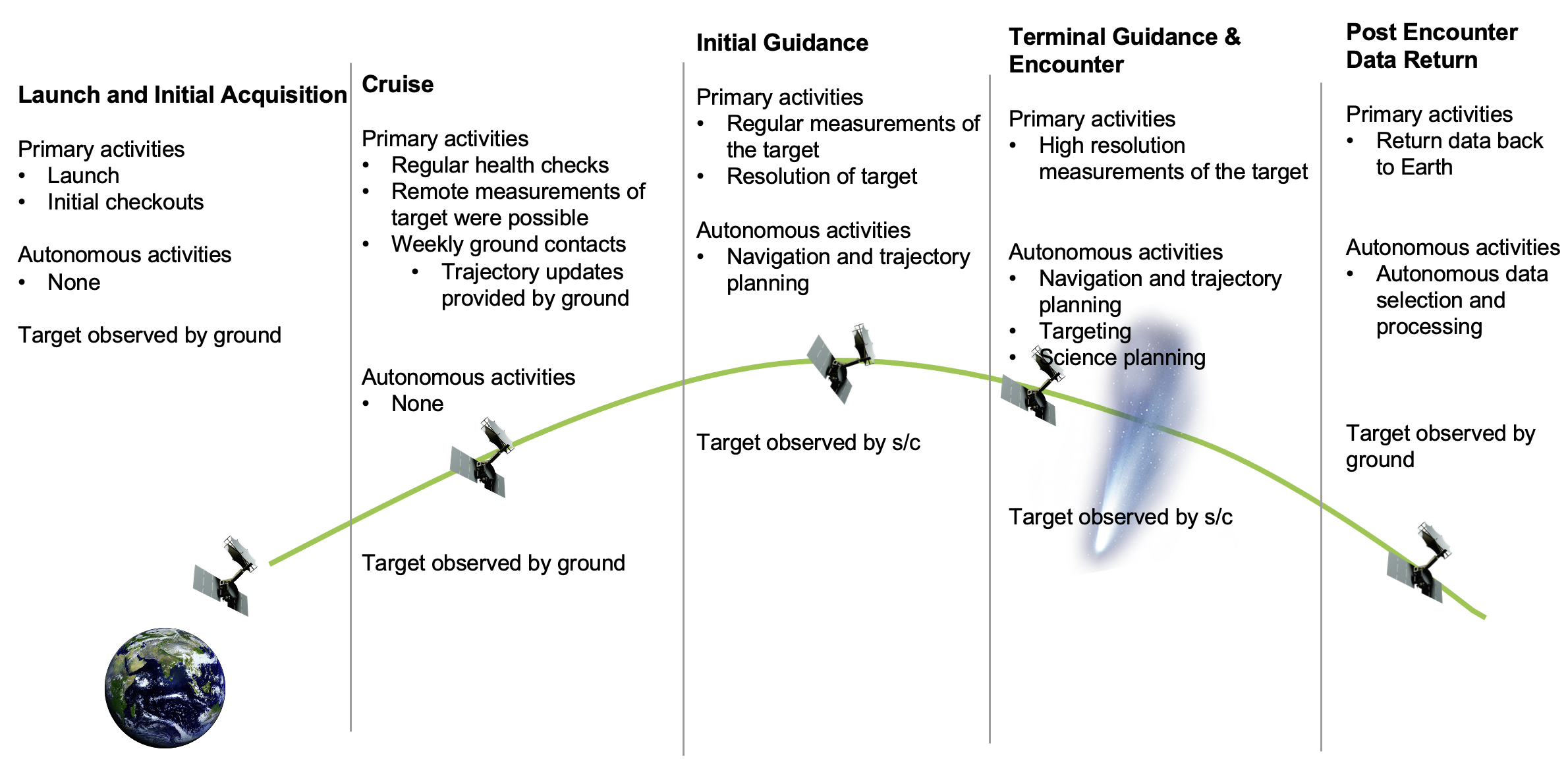}
    \caption{ISO Encounter Concept of Operations}
    \label{fig:conops}
\end{figure*}

Navigating to encounter an ISO requires state determination and trajectory correction to ensure a high-accuracy flyby. The journey to the ISO from launch to encounter is broken into three phases: Cruise, Approach, and Terminal Guidance. The primary phases are shown in Figure \ref{fig:conops}. Note that we also include initial guidance in that figure, which refers to the period of time when the ground can no longer command the spacecraft, but the autonomous navigation is not yet active.

During Cruise, navigators use two-way Doppler tracking to range the spacecraft and thus determine its orbit and trajectory. Ground-generated Trajectory Correction Maneuvers (TCMs) are used to put the spacecraft on the optimal trajectory to encounter the ISO. At the same time, space-based telescopic observations continue to refine the ISO trajectory. The process of orbit determination for both the ISO and the spacecraft is discussed in more detail in \cite{iso-2022}. During the cruise, the spacecraft is controlled using ground in the loop (GITL) and does not require a significant amount of autonomy. The novelty of this work is that, due to the rapid response nature of this mission concept, the knowledge of the ISO's trajectory is significantly reduced while the spacecraft is in flight. Ground-planned TCMs are used to correct the spacecraft trajectory to more accurately encounter the ISO as the knowledge of the ISO trajectory is improved over time.

The Approach phase starts when the first OpNav measurements are taken. This is typically when the OpNav measurements have a resolution greater than the resolution of Earth-based observatories, or when the target is detectable using the onboard OpNav imager. This start time can range from ten to dozens of days before the encounter and goes up until 24 hours from the encounter. In this phase, the ISO's trajectory is continuously refined using onboard OpNav, and GITL preparation for the encounter can start, including onboard instrument checkouts and other health checks. Autonomy onboard can begin to consider the trajectory planning by observing the relative state between the ISO and the spacecraft and planning the optimal control sequence to minimize encounter distance and resource consumption.

Terminal Guidance begins two hours out from the encounter. (This definition was used for consistency across simulations, but in a realistic mission scenario, we could consider the handoff to terminal guidance occurring when the knowledge of the encounter point is reduced to below about 100 km, which depends on the size and brightness of the target.) At this point, the spacecraft is acting fully autonomously. Though not studied in the scope of this work, science autonomy can also prepare the spacecraft to acquire the highest value data during close approach. During this phase, the spacecraft acquires the highest resolution images of the target and can precisely compute the relative location of the ISO. A form of autonomous navigation is used to minimize the distance between the spacecraft and the target at the time of encounter, known as "crossing the B-Plane." Some of the state of practice will be discussed later in this paper (see also \cite{autonav}).

\subsection{Optical Navigation}

Optical navigation, or OpNav, is the process of determining the relative state between a target body and a spacecraft using images obtained by onboard cameras. OpNav has proved to be a highly accurate means to determine state and is frequently used as a primary means of state and orbit determination as the spacecraft camera starts resolving the target \cite{OpNav}. Optical navigation works by finding the center of the brightness of the observed target and the centers of cataloged stars. In general, the accuracy of the method is proportional to the pointing knowledge error plus the metric size of the pixel, meaning that higher resolution cameras are sought to increase the OpNav quality, especially when dealing with small targets. For this study, the authors assumed a camera with an iFOV of 18 $\mu$rad based on the Advanced Pointing Imaging Camera narrow-angled camera (NAC) developed at JPL \cite{apic}, and attitude knowledge of 10 $\mu$rad from the wide-angle camera (WAC). The simulated OpNav measurements are used as the basis of knowledge for the autonomous navigation phase of the trajectory. In this work, both the state of practice autonomous navigation and the learning-based G\&C utilize the same basis of knowledge. In reality, there will be some coupling effects between maneuver execution and knowledge as things such as maneuver execution error increase the amount of knowledge uncertainty. Assessment of those coupling effects is not being considered in this study and is reserved for future work.

\subsection{Ground in the Loop Planning}

Traditional small-body flybys are operated and commanded using GITL. The spacecraft will downlink OpNav images, which navigators on the ground process and use to plan and perform a set of maneuvers that maximize delivery accuracy and update the knowledge of where and when the flyby with occur. GITL planning can be very effective and makes use of substantial processing power available on the ground, but it requires a significant amount of time to execute. Ground network availability, radio transmission time, light time, data processing, and analysis time all add to the total time it takes for the spacecraft to execute new commands using recently acquired OpNav measurements. While included here for completeness, ground network availability is generally not an issue for mission-critical events. Typically, GITL planning is based on OpNavs taken from 36 to 24 hours out from the close approach, but there remains too much state uncertainty in the case of ISOs to build a plan \cite{autonav} accurately.

For this analysis, the authors assumed that the spacecraft would operate in an autonomous mode for the final 2 hours prior to the close approach. Prior to that time, some GITL planning is used to correct the spacecraft trajectory and deliver the spacecraft to the autonomous phase. The spacecraft's initial position error is assumed to be equal to our knowledge uncertainty at the handoff to autonomy, which is based on ground-based observations. After handoff to autonomy, the spacecraft is assumed to be able to resolve the relative state using onboard autonomous optical navigation via AutoNav. 

\subsection{AutoNav}

AutoNav is a proven technology that was first used on the mission Deep Space 1, and has since operated during flybys Borrelly, Wild 2, Tempel 1, Hartley 2, and Annefrank \cite{autonav_ds1}. It enables a spacecraft to autonomously image a target, register those images to derive a measurement, and use that measurement to update the spacecraft's relative state knowledge. The knowledge can then be used to design a maneuver for more precise delivery, or the knowledge can be used for precise pointing of science instruments during encounters.

\subsection{Covariance Analysis}

To characterize the relative uncertainties during the autonomous terminal phase, the authors utilize JPL's Mission Analysis, Operations, and Navigation Toolkit Environment (MONTE) software to perform covariance analysis for each spacecraft and ISO trajectory pair. The details of this analysis are covered in \cite{ISO_covariance_study}. In summary, each ISO's uncertainty over time is estimated via simulated ground-based astrometric observations with realistic constraints based on view geometries. During the cruise, the spacecraft's uncertainty over time is estimated via Earth-based radiometric tracking data. Then, during the approach, the effects of OpNav data are added and drive down the ISO's uncertainty and overall relative uncertainty up until the final GITL cutoff. After this, the autonomous phase is modeled by removing radiometric tracking and altering the OpNav cadence to resemble more rapid autonomous tracking. With this method, the authors are able to, with high fidelity, estimate the entire relative uncertainty profile through the mission all the way to the encounter. This data can then be used to inform control design. 

\section{Deep Learning-based Guidance and Control}
The high velocity and state uncertainty challenges in encountering ISO have led us to investigate the emerging areas of learning-based control, where the theoretical performance guarantees of control theory are combined to enhance autonomous Guidance and Control (G\&C) with conventional black-box AI-based techniques~\cite{tutorial}. In \cite{tsukamoto2022}, the authors at Caltech and JPL collaborated to investigate Neural-Rendezvous, a novel deep learning-based G\&C framework for encountering fast-moving targets, including ISOs, robustly, accurately, and autonomously in real-time even with the limited computational capacity of current spacecraft. It uses the spectrally-normalized deep neural network (SN-DNN) to approximate model predictive control (MPC), which is known to be near-optimal in terms of dynamic regret, \ie{}, the MPC performance minus the optimal performance in hindsight~\cite{NEURIPS2020_155fa095} with its optimality measured by the total $\Delta V$. The strength of Neural-Rendezvous lies in its control theoretical stability, robustness, and optimality guarantees even with the use of machine learning, while eliminating the need for substantial onboard computation in extensively utilizing the onboard state estimates obtained by the aforementioned optical navigation.

\subsection{Development of Algorithm}
Consistent with the objective of encountering ISOs, Neural-Rendezvous trains an SN-DNN-based guidance policy with a novel loss function for directly imitating the MPC state trajectory, performing dynamics integration. It is then augmented by learning-based min-norm feedback control, which provides an optimal and robust control input, expressed in an analytical form, that minimizes its instantaneous deviation from that of the SN-DNN guidance policy under the incremental stability condition as in the one of contraction theory~\cite{tutorial}.

In particular, Neural-Rendezvous is shown to have the following performance guarantees: 1) a verifiable optimality gap with respect to the MPC policy; and 2) formal state tracking error bound with respect to the desired state trajectory, which decreases exponentially in expectation with a finite probability, robustly against the state uncertainty. The latter indicates that the terminal spacecraft deliver error at the ISO encounter is probabilistically bounded in expectation, where its size can be modified according to the mission requirement by tuning its G\&C parameters. 

\subsection{Testing in Caltech Software and Hardware Environment}
The Caltech MATLAB-based software simulation setup is developed for testing the performance of Neural-Rendezvous and other autonomous GNC algorithms in exploring ISOs, where its details can are discussed in~\cite{tsukamoto2022} and the results are partially discussed also in~Section~\ref{sec:comparison}. A YouTube video which visualizes the software simulation results can be found {\color{caltechgreen}\underline{\href{https://youtu.be/4KPaqSpFMEU}{here}}}.

\begin{figure}
    \centering
    \includegraphics[width=.4\textwidth]{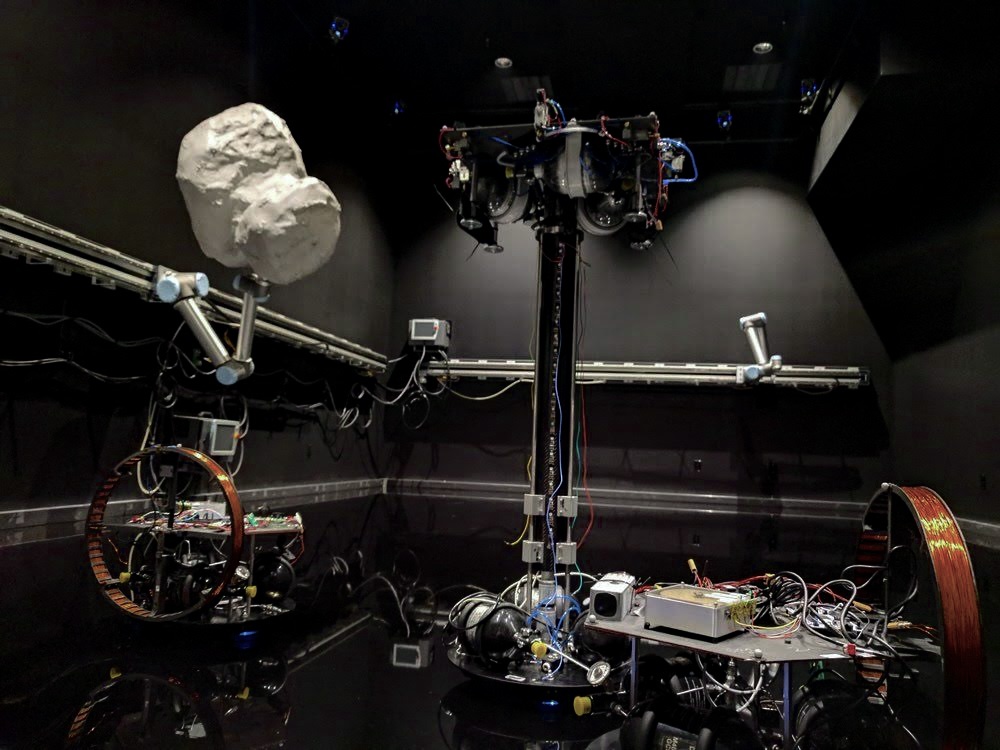}
    \caption{M-STAR spacecraft at Caltech.}
    \label{fig:sc_simulator}
\end{figure}

There is ongoing work on validating Neural-Rendezvous's performance on an actual spacecraft using Caltech's six-degree-of-freedom robotic spacecraft simulator (M-STAR)~\cite{SCsimulator} shown in Figure~\ref{fig:sc_simulator}, which runs on NVIDIA Jetson TX2 as the main computer to run the GNC and perception algorithms. It is controlled by eight thrusters at a 2 Hz control frequency, with the motion capture camera system running at 100 Hz. The hardware simulation is performed by scaling down the ISO dynamics to that of the S/C simulators, with the uncertainty assumed to be the ones given by AutoNav. See~\cite{SCsimulator} to learn more about the simulator specifications. The preliminary results can be found~{\color{caltechgreen}\underline{\href{https://drive.google.com/drive/folders/1Aw4Mr5tG1uf-YRwAIvG3pGBNSc626yvd?usp=sharing}{here}}}.

\section{Testing Neural-Rendezvous in the JPL SmallSat Dynamics Testbed}
The JPL SmallSat Dynamics Testbed (SSDT) is a high-fidelity dynamics simulation that models uncertainties and constraints involved in spaceflight. JPL has a hardware simulation component to the SSDT, but this testing used only the software version of the testbed, which uses Simulink. The purpose of the SSDT for this investigation is to evaluate the performance of G\&C techniques for ISO exploration when implemented on a space mission. The SSDT enables a trade analysis and sensitivity study on the impacts of uncertainties and design variables on the algorithm's performance during terminal guidance.

\subsection{SSDT Structure and Input Parameters}
This particular study used only 3-DOF translation rather than including attitude dynamics because we are not interested in maneuvering fuel consumption yet. The SSDT imports trajectory data from synthetic ISOs and an ideal spacecraft trajectory from launch to target intercept, and uses this to initialize the spacecraft and ISO states at two hours prior to encounter. The SSDT then propagates the states using a two-body inertial simulation.

The SSDT parameterizes the uncertainties and design variables to facilitate a sensitivity analysis against their variations. The parameters of interest are the following:

\begin{itemize}
\item Relative position knowledge uncertainty
\item Thruster uncertainty
\item Thruster resolution
\item True delivery error
\item Trajectory
\end{itemize}

The relative position uncertainty feature simulates the optical navigation sensor by overlaying time-varying, zero-mean Gaussian noise dictated by the covariance evolution model generated from MONTE onto the true state as propagated by the simulation. The thruster uncertainty was also a zero-mean Gaussian distribution but time-invariant. The thruster resolution simulates a noncontinuous range of force values from the thruster. The delivery error refers to the position and velocity error from the ideal trajectory at the start of autonomous navigation, which would be caused by uncertainty when executing the last GITL trajectory correction maneuver. This error was set as the one-sigma uncertainty value of the uncertainty model at the start of optical navigation. For a small number of runs, a delivery error was randomized to observe its distribution's effect on the distribution of the performance, especially for miss distance. Lastly, these variables were tested across different trajectories. There were 207 model trajectories run in the SSDT that have varying spacecraft-ISO relative velocities and inclinations. Only a subset of these were run across variations in the other parameters in the interest of time.

\subsection{Choosing the Performance Variables}
The performance variables of interest were the following:
\begin{itemize}
\item ISO miss distance (relative SC-ISO distance at time of closest approach)
\item Delta V and fuel used
\item Computational load
\end{itemize}
The delta V and fuel used were calculated based on a 150 kg spacecraft only thrusting in translation. Computational load is of interest due to the compute and power cost differences across different guidance and control algorithms. The SSDT gives a basic relative measurement of this by timing the algorithm run time per control cycle and then summing the run times throughout the flight.

\subsection{Designing the Sensitivity Analysis}
We want the input parameters to span values that may be seen in a trade study during spacecraft development to understand any possible drawbacks of the algorithm.
It's not necessary to run every variation across the full trajectory space, so a deeper sensitivity analysis was done on a smaller set of trajectories and can be extrapolated to the other trajectories. The selected variations can be found in Table \ref{tab:ssdt-params}. A coefficient was used as the controllable parameter for relative position uncertainty, which is a factor that magnifies the Gaussian noise (a coefficient of 0.5 means less uncertainty). Test cases 1 and 2 intend to capture the performance spread with and without delivery error uncertainty (constant versus noisy delivery error across several runs). Test cases 3-6 are intended to capture sensitivities to changing parameters individually. Since each run had a random distribution, each set of parameters was run at least 3 times to capture a signal-to-noise of the performance's sensitivity to the parameter. The last test case ran repeated runs of a best-case and a worst-case scenario.

\begin{table*}[]
\caption{Variations on input parameters were broken into several test cases to focus on individual impacts on performance.}
\label{tab:ssdt-params}
\begin{tabular}{|l|l|l|l|l|l|l|l|}
\hline
Input Parameter                                                                             & Test Case 1 & Test Case 2 & Test Case 3 & Test Case 4 & Test Case 5                                                 & Test Case 6 & Test Case 7                                            \\ \hline
\begin{tabular}[c]{@{}l@{}}Relative Position \\ Knowledge Coefficient\end{tabular}          & 1           & 1           & 0.5, 1, 3   & 1           & 1                                                           & 1           & 0.5/3                                                  \\ \hline
\begin{tabular}[c]{@{}l@{}}Thrust Uncertainty \\ Coefficient\end{tabular}                   & 1           & 1           & 1           & 0, 10, 100  & 1                                                           & 1           & 0.5/3                                                  \\ \hline
\begin{tabular}[c]{@{}l@{}}Thruster \\ Resolution (N)\end{tabular}                          & 0.01        & 0.01        & 0.01        & 0.01        & \begin{tabular}[c]{@{}l@{}}0.01, 0.1,\\  1, 10\end{tabular} & 0.01        & 0/0.1                                                  \\ \hline
\begin{tabular}[c]{@{}l@{}}Delivery Position and \\ Velocity Error Coefficient\end{tabular} & 1           & 1           & 1           & 1           & 1                                                           & 0, 1, 10    & 0.1/10                                                 \\ \hline
\begin{tabular}[c]{@{}l@{}}Delivery Error \\ Uncertainty?\end{tabular}                      & False       & True        & False       & False       & False                                                       & False       & \begin{tabular}[c]{@{}l@{}}False/\\ False\end{tabular} \\ \hline
\end{tabular}
\end{table*}

\subsection{Results and Work To Go}
The final results of Neural-Rendezvous testing in the SSDT are still in work. Miss distance values did not correlate closely with the Caltech and MONTE simulations to fully validate the results from the SSDT simulation. However, initial runs of the relative position knowledge uncertainty trade indicated a large effect of down-track position knowledge noise on cross-track miss distance with the neural-rendezvous technique, despite having good cross-track knowledge from the optical navigation sensor. For problems like these, a simple baseline controller was designed to run in the SSDT to debug and validate the sim and provide a comparison baseline to other controllers. The simple guidance scheme with PID control seemed to do better with miss distance throughout the sensitivity study, so future work involves ensuring the N-R algorithm is integrated properly into the simulation. After this is done, this study using the SSDT could involve other parameters such as control frequency and control authority.

\section{Comparison Between State of Practice and Neural-Rendezvous}
\label{sec:comparison}
\begin{figure}
    \centering
    \includegraphics[width=0.45\textwidth]{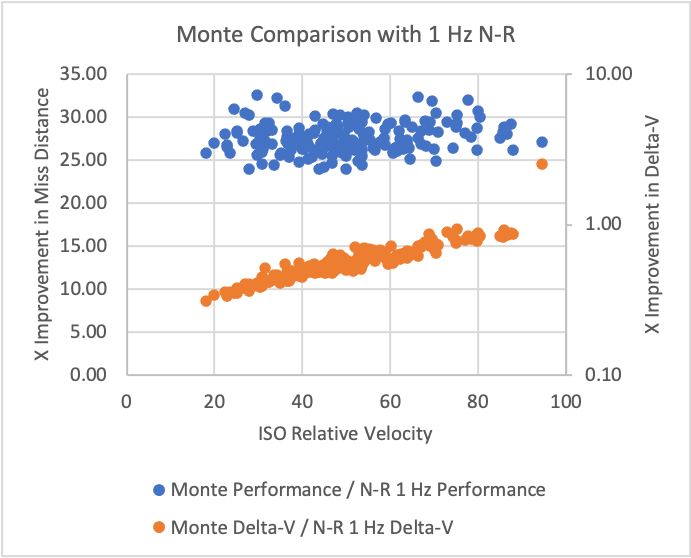}
    \caption{N-R algorithm achieves 25-30x better performance at the cost of ~4x more $\Delta V$ as compared to AutoNav}
    \label{fig:relative-comparison}
\end{figure}

\begin{figure*}
    \centering
    \subfloat{
        \includegraphics[width=0.45\textwidth]{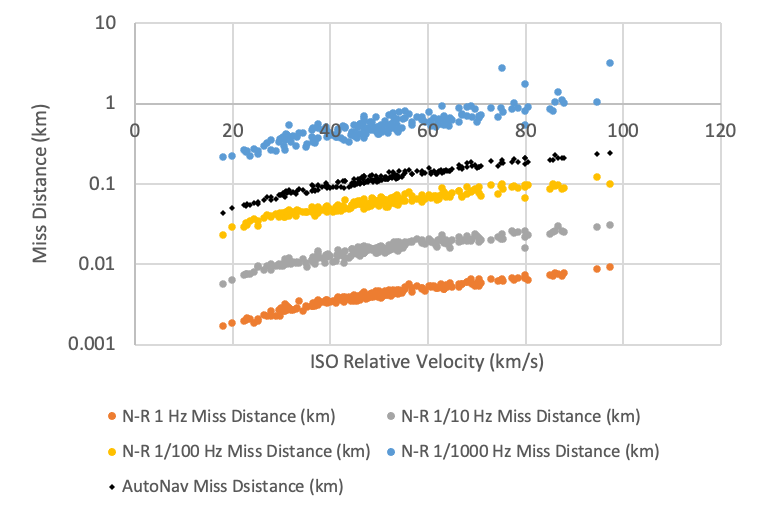}
        }
    \subfloat{
        \includegraphics[width=0.45\textwidth]{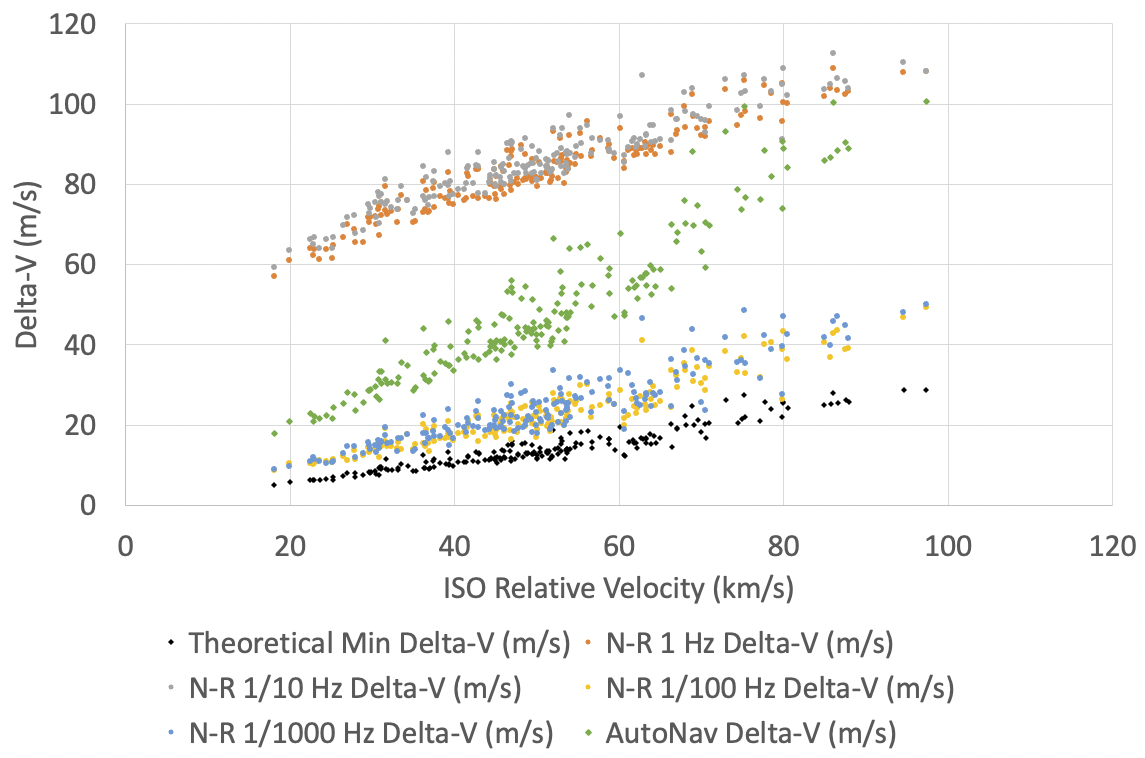}
        }
    \caption{N-R algorithm performance degrades with decreasing control frequency. Left: Absolute miss distance as a function of ISO relative velocity. Right: $\Delta V$ consumption as a function of relative velocity.~}
    \label{fig:absolute-comparison}
\end{figure*}

The novel G\&C approach regularly performs up to 25-30$\times$ better in terms of miss distance than the traditional AutoNav approach as assessed using MONTE at the expense of higher $\Delta V$ consumption. Figure \ref{fig:relative-comparison} shows the relative performance of the novel G\&C approach compared to traditional AutoNav. Blue dots correspond with X times the improvement of the novel G\&C approach over AutoNav on the right axis (e.g., 25 corresponds to a 25X improvement in miss distance). Orange dots correspond with X times difference in $\Delta V$ consumption. A value less than 0 indicates a reduction in performance (e.g., 0.25 means 4X more $\Delta V$). The significant improvement of the novel approach is likely a result of both increased control authority and better-optimized maneuvers. Figure \ref{fig:absolute-comparison} shows the absolute performance and resource consumption of both the traditional AutoNav approached simulated in MONTE and the novel G\&C approach simulated in the Caltech software simulator.

The performance of the N-R algorithm is sensitive to the control frequency, but optimizing maneuver placement can significantly improve the performance of the algorithm at lower control frequencies. Presently, the algorithm equally spaces maneuvers, meaning that the final maneuver may be placed well before the close approach. With the current maneuver planning schedule, the N-R algorithm's performance drops to below that of AutoNav at 1/1000 Hz, as shown in Figure \ref{fig:absolute-comparison}. Optimization of maneuver placement may significantly enhance the results and present an improved performance. This is an area for future work in the development of the N-R algorithm for use in high-velocity small bodies flybys.

To compare the performance of an algorithm, we introduce a term, Navigation Efficiency, $\eta_\mathrm{nav}$, which is maximum at very low miss distances and $\Delta V$, and minimum at large miss distances and $\Delta V$. This term can describe the performance of a G\&C algorithm, whose purpose is to minimize miss distance with minimal $\Delta V$.

\begin{equation}
    \eta_\mathrm{nav} = \frac{1}{\Delta V \times dist_{miss}}
\end{equation}

\begin{figure}
    \centering
    \includegraphics[width=0.45\textwidth]{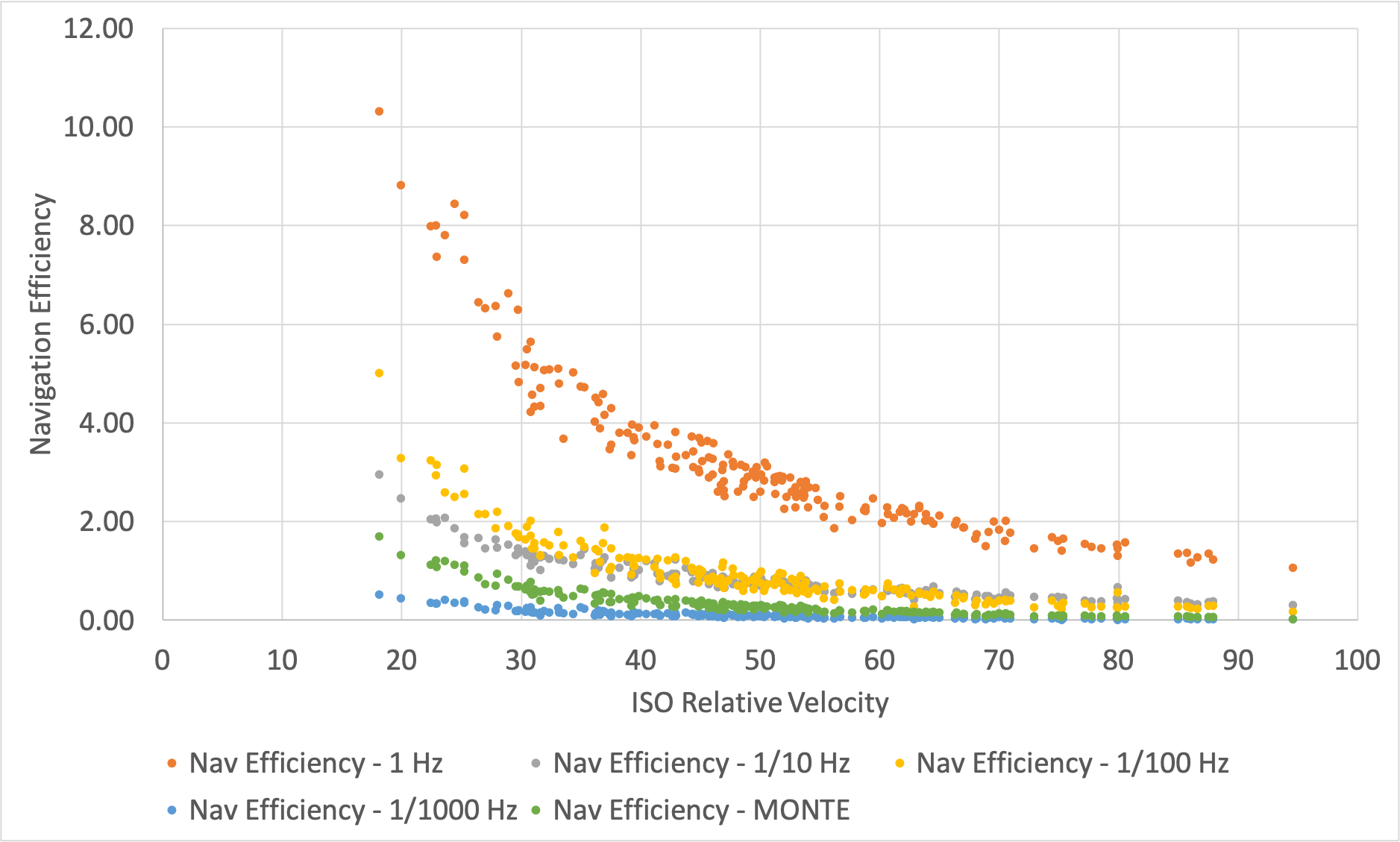}
    \caption{Navigation efficiency decreases with reduced control frequency and increased relative velocity.}
    \label{fig:nav_eff}
\end{figure}

The N-R G\&C approach provides a very efficient algorithm with high $\eta_\mathrm{nav}$ at high control frequencies. 
However, as also described above, the performance degrades significantly with reduced control frequency. We also see that as our flyby increases in relative velocity, effectively making the flyby more challenging, $\eta_\mathrm{nav}$ decreases. 
In Figure \ref{fig:nav_eff}, the green dots represent the JPL State of Practice AutoNav results. 
Again, the performance of the N-R G\&C algorithm drops below that of AutoNav at less than 1/100 Hz. Higher frequency N-R results in much higher $\eta_\mathrm{nav}$ than the current state of practice, emphasizing the strong performance of the novel approach under ideal conditions.

\section{Conclusions}

Performing a precision flyby or impact of a high-velocity interstellar object is challenging. The extremely high relative velocities make ground in the loop effectively impossible to use as a navigation approach, and even makes traditional autonomous navigation techniques sub-optimal. To deal with the highly dynamic environment, we developed and present Neural-Rendezvous guidance and control technique, which we demonstrate to be extremely efficient, meaning high delivery accuracy for relatively lower $\Delta V$: a 30x improvement in delivery accuracy over the state of practice AutoNav at the cost of about 4x more $\Delta V$, up to 100 m/s total (see Figure \ref{fig:absolute-comparison}). However, the N-R approach requires a very high control frequency which could pose a challenge for both onboard computing and thruster control authority based on current commonly used technology. As the allowable control frequency of the N-R decreases, its performance similarly decreases. Below a control frequency of 1/100 Hz, the N-R algorithm begins to perform worse than AutoNav using these initial assumptions. Future work to analyze maneuver placement and re-optimize the algorithm for lower control frequencies may recuperate some of the degraded performance and confirm that the N-R approach is also more efficient in cases with reduced control frequency.

\section{Future Work}

The work described in this paper presents an initial development and assessment of a critical technology required for a future ISO flyby. There remains work to go both in the development of ISO flyby architectures and in the development of this and future autonomous G\&C approaches. Future work described below is broken into two sections: additional work required to enable an ISO encounter and additional work on the autonomous G\&C approach.  

\subsection{Additional Work to Enable ISO Encounter}

Encountering an ISO requires a paradigm shift from the common way of formulating, implementing, and operating a mission. Science missions traditionally visit targets with accurately known ephemerides that result from decades of observations of a celestial body. While NASA has considered and launched missions that result from unique opportunities such as planetary alignments (e.g., Voyager \cite{voyager}, Trident \cite{trident}), most missions are not strictly driven by unique launch windows. ESA is also funding the Comet Interceptor mission, which will be a first-of-its-kind mission to encounter a recently-discovered Long Period Comet \cite{interceptor} but will likely not have enough onboard delta-V to flyby an ISO \cite{interceptor-2}. An ISO mission, on the other hand, would require \textit{rapid response}, launch from Earth or a staging orbit shortly after initial detection, while the target's orbit is still poorly constrained, and on an extremely tight schedule. 

In order to enable an ISO mission, the mission community needs to continue investment and exploration in rapid-response architectures that can enable a mission to go from initial detection of an ISO to launch, either from Earth or from a staging orbit, within months to years. Specifically, the authors recommend the following follow-on activities to assess rapid response capability:

\begin{itemize}
    \item End-to-end simulation of detection, characterization, launch, and operations
    \item Development of an agile spacecraft that can be rapidly integrated and tested 
    \item Assessment of in-space vs. on-ground storage vs. build-after-launch architectures for rapid response
    \item Continued development of synthetic ISO databases using high-fidelity galactic gravity models
    \item Continued assessment of presumed ISO detection capabilities with next-generation observatories including, Vera C. Rubin
\end{itemize}

Many of these items are also directly applicable to rapid response to visit long-period comets, which often are also discovered close to perihelion or for planetary defense. 

\subsection{Additional Work on Autonomous G\&C}

This body of work describes only an initial assessment of a novel G\&C performance algorithm as compared to AutoNav. The state of practice AutoNav has proven successful in-flight on Deep Impact and is well validated. The novel G\&C technique shows promise but needs to undergo a similar level of validation to ensure that it is robust to off-nominal scenarios, that it can be applicable to targets that are unlike those in the training set, and that it can still be useful with a reduced control frequency. We also recommend several activities to improve the performance of the novel G\&C controller and possibly improvements to the state of practice AutoNav.

\begin{itemize}
    \item Optimize maneuver placement in both the novel N-R approach and AutoNav
    \item Test the N-R approach under off-nominal conditions, including worse-than-expected thruster performance, greater handoff error, or worse knowledge of the target
    \item Test the N-R approach using different ISO trajectories in the training set and using a training and test set generated using different ISO models
    \item Integrate knowledge acquisition with improved control in the N-R approach
    \item Assess statistical burn time optimization given estimated uncertainty over time
    \item Multi-agent setup that could enable more aggressive ISO exploration objectives
\end{itemize}

\appendices{}              

\acknowledgments
Part of the research was carried out at the Jet Propulsion Laboratory, California Institute of Technology, under a contract with the National Aeronautics and Space Administration. Government sponsorship acknowledged.

\bibliographystyle{ieeetr}
\bibliography{bib.bib}

\thebiography
\begin{biographywithpic}
{Benjamin P. S. Donitz}{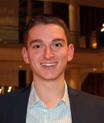} received his Bachelor's and master's degrees in Aerospace and Space Engineering, respectively, from The University of Michigan. He now works at the Jet Propulsion Laboratory in the Project System Engineering and Formulation section where he is involved with the development of early-phase mission concepts. Since joining JPL in 2019, Benjamin has worked on several mission concepts and studies evaluating the feasibility of sending dedicated spacecraft to interstellar objects and long-period comets. He also works on a variety of other mission and mission concepts, including Mars Sample Return, and with JPL's concurrent design and engineering teams, Team-X and A-Team.
\end{biographywithpic} 

\begin{biographywithpic}
{Declan Mages}{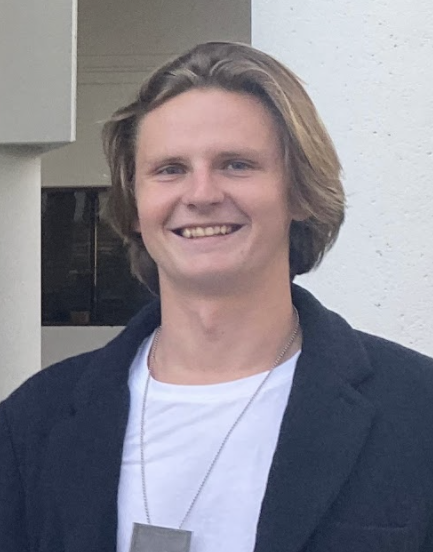} is a navigation engineer in JPL's outer planet navigation group, specializing in optical navigation. He graduated with a Master's in Aerospace Engineering from Cal Poly San Luis Obispo. His thesis focused on characterizing the challenges of navigating encounters with interstellar objects. Since arriving at JPL, he has worked as an optical navigator on the New Horizons encounter with Arrokoth and is now preparing to navigate the Double Asteroid Redirection Test while researching new optical navigation processing techniques.
\end{biographywithpic}

\begin{biographywithpic}
{Hiroyasu Tsukamoto}{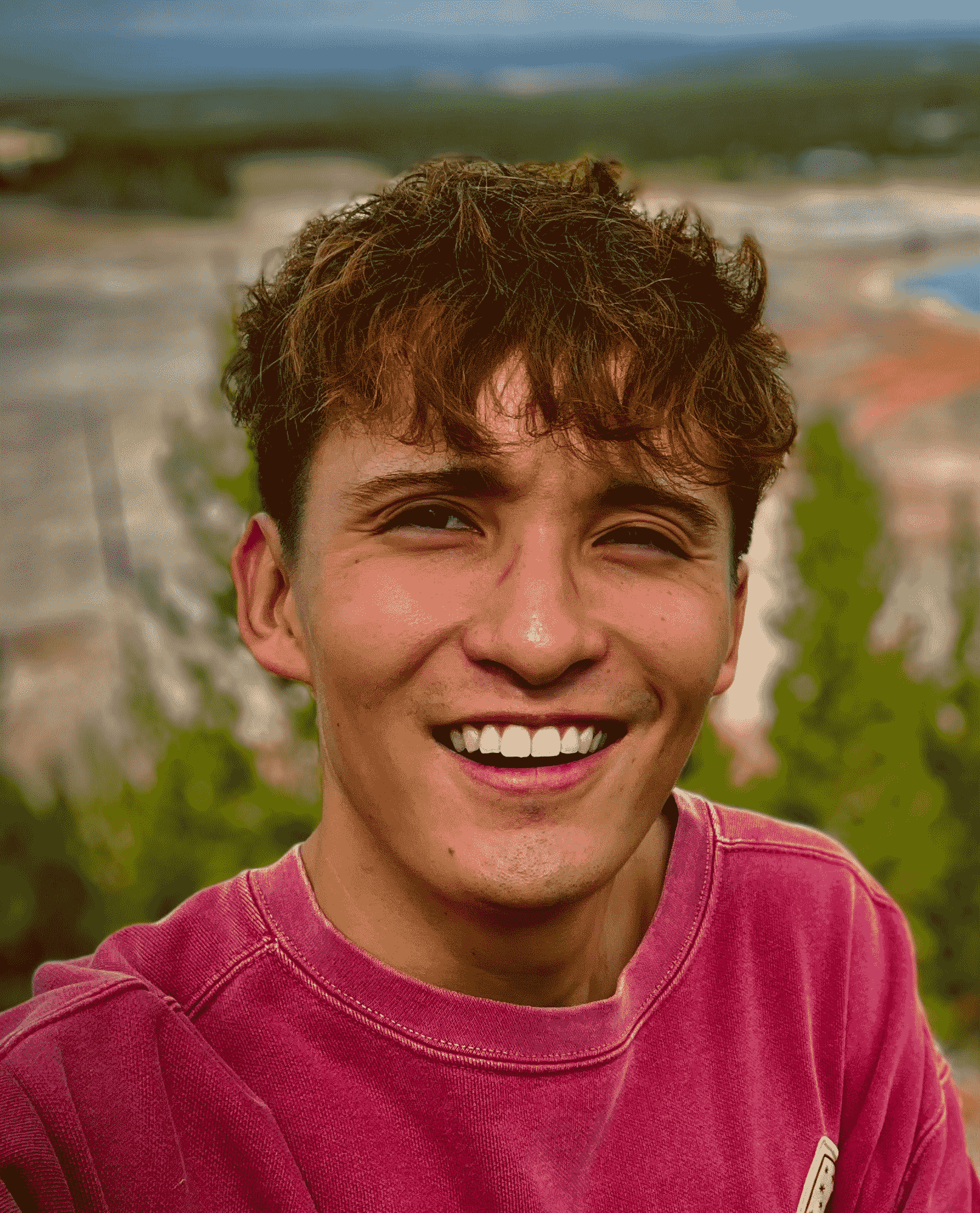} is an Assistant Professor of Aerospace Engineering at the University of Illinois at Urbana-Champaign and the director of the ACXIS Laboratory (Autonomous Control, Exploration, Intelligence, and Systems). Prior to joining Illinois, he was a Postdoctoral Research Affiliate in Robotics at the NASA Jet Propulsion Laboratory, where he contributed to the Science-Infused Spacecraft Autonomy for Interstellar Object Exploration and Multi-Spacecraft Autonomy Technology Demonstration projects. He received his Ph.D. and M.S. in Space Engineering (Autonomous Robotics and Control) from Caltech in 2018 and 2023, respectively, and his B.S. degree in Aeronautics and Astronautics from Kyoto University, Japan, in 2017. He is the recipient of several awards, including the William F. Ballhaus Prize for the Best Doctoral Dissertation in Space Engineering at Caltech and the Innovators Under 35 Japan Award from MIT Technology Review. More info: \myhref{https://hirotsukamoto.com}{https://hirotsukamoto.com}.
\end{biographywithpic}

\begin{biographywithpic}
{Peter Dixon}{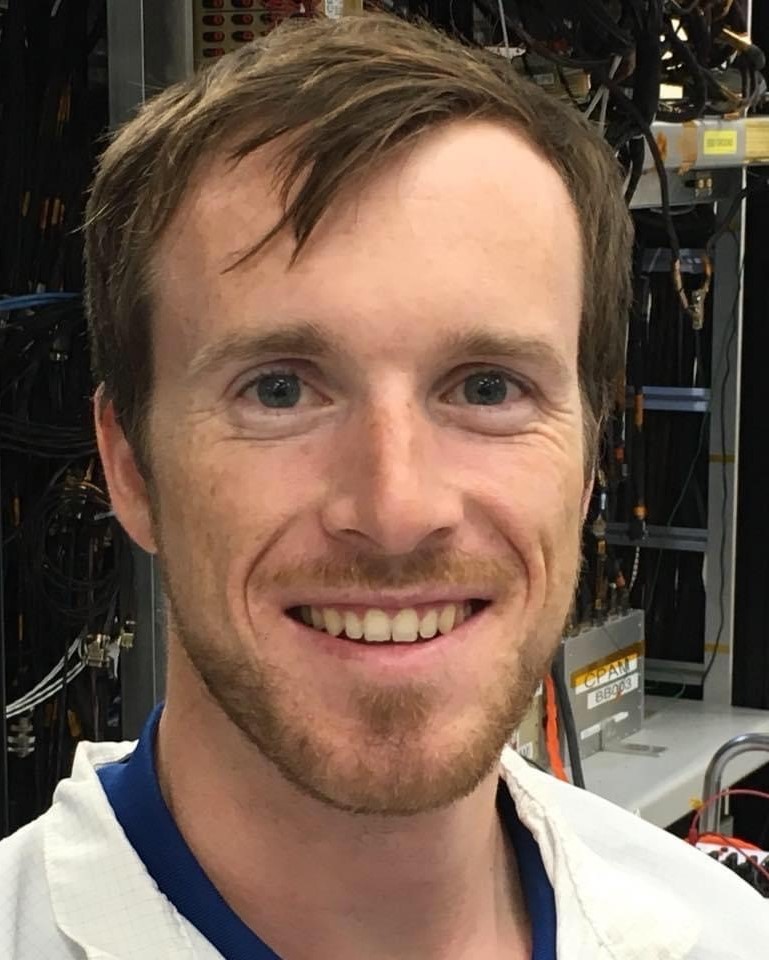} is a G\&C systems engineer at the Jet Propulsion Laboratory with a background in avionics. He has worked on verification and validation for Mars 2020, Europa Clipper, and Psyche avionics for most of his career at JPL, with a specialization in onboard clocks and timekeeping. His interests lie in G\&C systems engineering and early phase G\&C subsystem design. He has recently started on Mars Sample Return's Sample Retrieval Helicopter Aerial Mobility in early design work. Peter graduated from Case Western Reserve University with a B.S. in Mechanical and Aerospace Engineering and from The University of Michigan with a Master's in Space Systems Engineering.
\end{biographywithpic}

\begin{biographywithpic}
{Damon Landau}{biopics/D_landau_photo_10x8} is a systems engineer at the Jet Propulsion Laboratory, where his primary interests are mission formulation and trajectory optimization. He routinely leads the technical development of proposals for NASA's EVM, MIDEX, SIMPLEx, Discovery, and New Frontiers programs. Damon formulated the initial trajectory concepts for the Discovery mission to Psyche, the Near-Earth Asteroid Scout sailcraft, and the twin MarCO relay cubesats. He received his B.S., M.S., and PhD in Aeronautics and Astronautics from Purdue University.
\end{biographywithpic}

\begin{biographywithpic}
{Soon-Jo Chung}{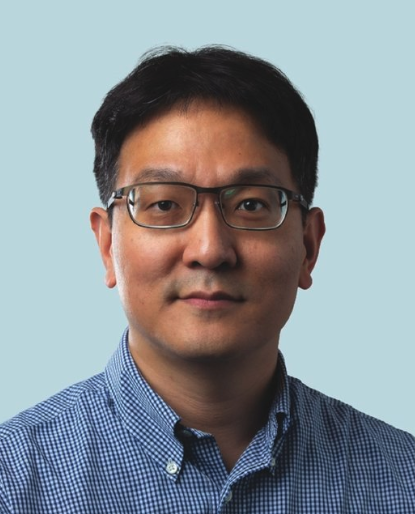} received the B.S.degree (\textit{summa cum laude}) in aerospace engineering from the Korea Advanced Institute of Science and Technology, Daejeon, South Korea, in 1998, and the S.M. degree in aeronautics and
astronautics and the Sc.D. degree in estimation and control from Massachusetts Institute of Technology, Cambridge, MA, USA, in 2002 and 2007, respectively. He is currently the Bren Professor of Aerospace and Control and Dynamical Systems and a Jet Propulsion Laboratory Senior Research Scientist at the California Institute of Technology, Pasadena, CA, USA. He was with the faculty of the University of Illinois at Urbana-Champaign during 2009–2016. His research interests include spacecraft and aerial swarms and autonomous aerospace systems, and in particular, on the theory and application of complex nonlinear dynamics, control, estimation, guidance, and navigation of autonomous space and air vehicles.
\end{biographywithpic}

\begin{biographywithpic}
{Erica Bufanda}{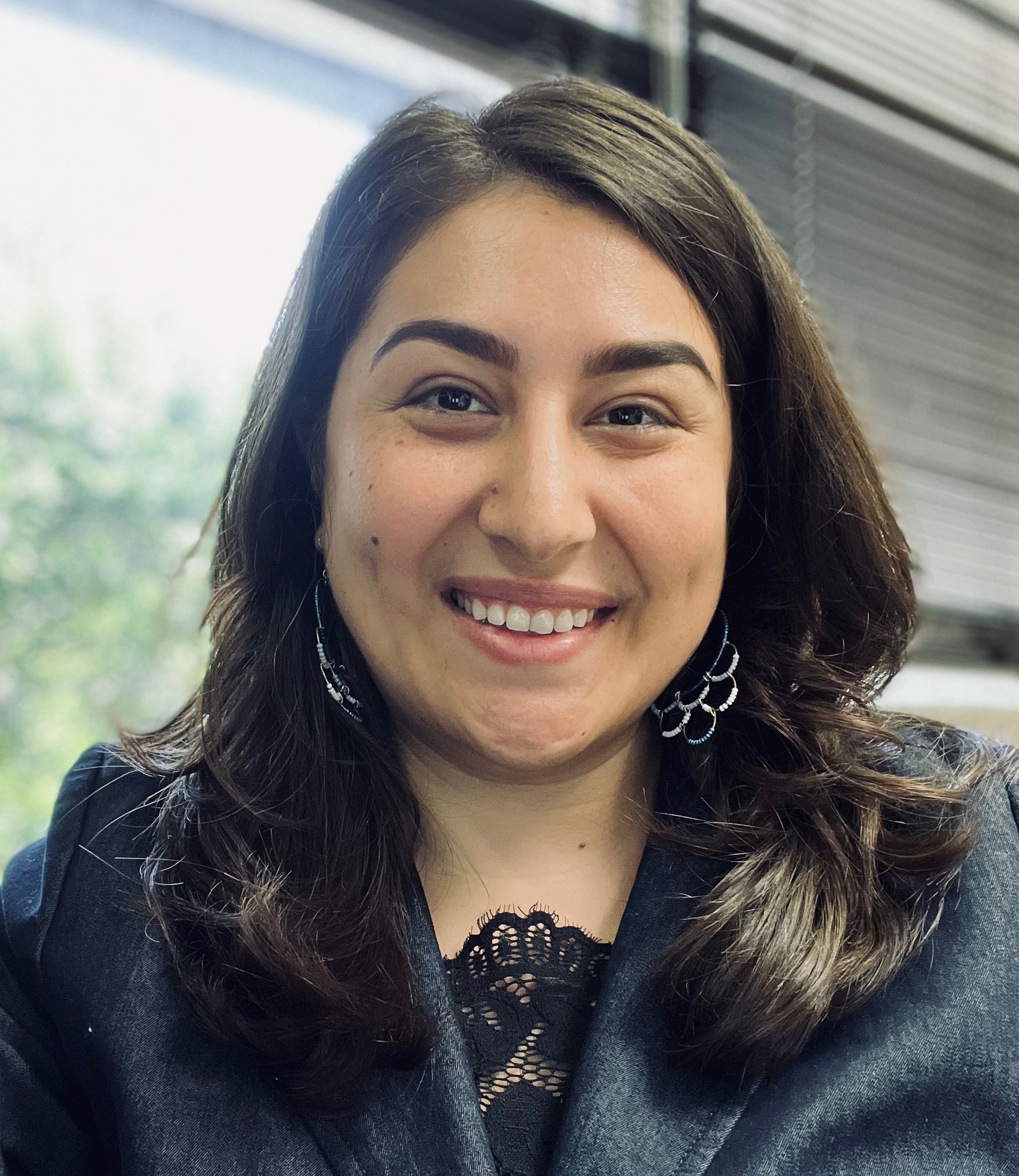} is an Astronomy PhD candidate at the University of Hawaii, Manoa. Her thesis is on characterizing a large sample (90+) of comets from the Oort Cloud. To do this, she is the P.I. of two observing programs utilizing professional telescopes CFHT and Gemini North on Maunakea. With her data, she constrains comet orbits, assesses comae and nuclei colors, and models cometary sublimation and dust. Her interests and goals are to use comet populations to answer fundamental questions in planetary science and provide a diverse ground-based data set to inform future spacecraft missions about the types of environments we might encounter.
\end{biographywithpic}

\begin{biographywithpic}
{Michel Ingham}{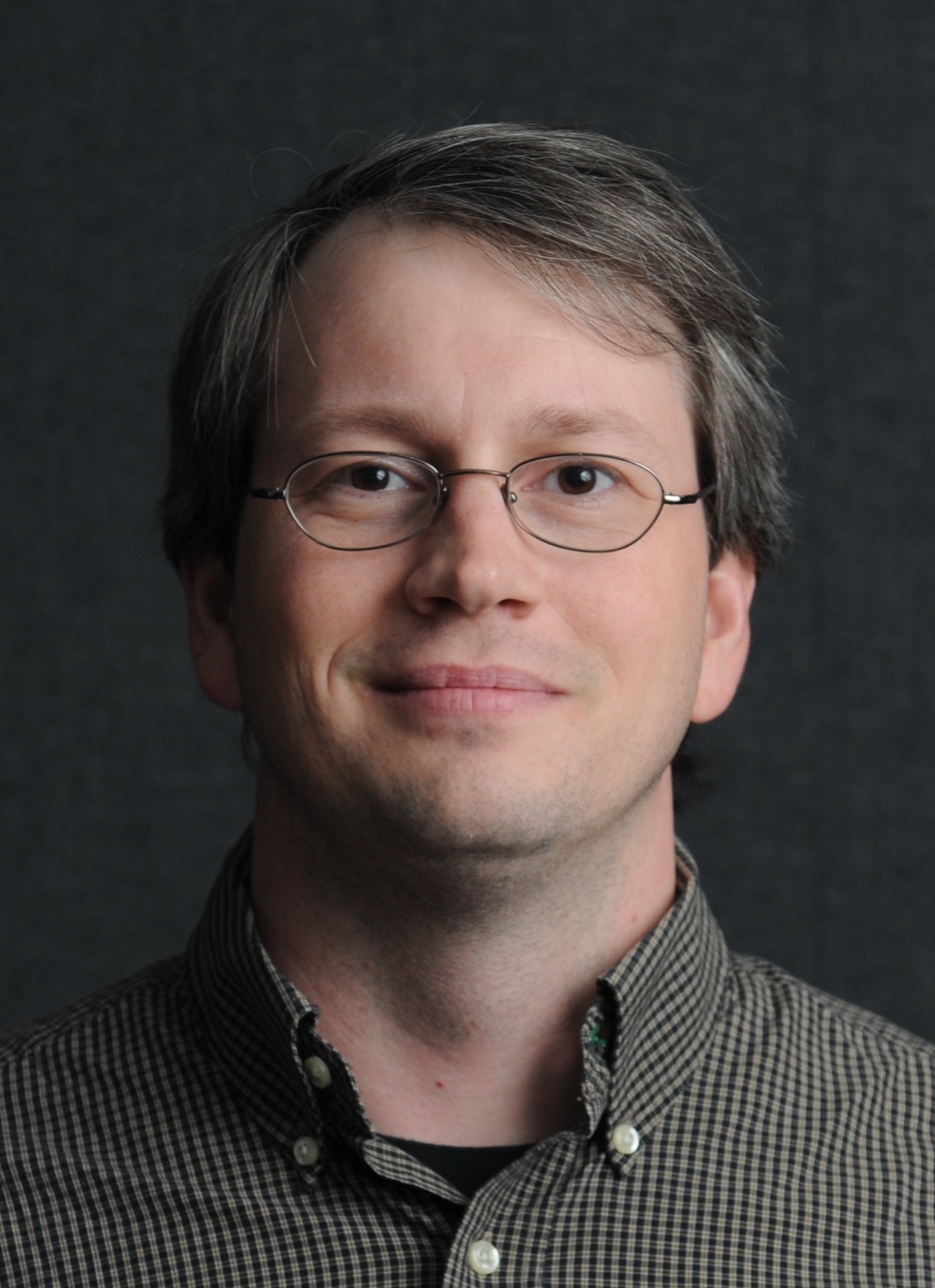} is the Chief Technologist of JPL's Systems Engineering Division, responsible for spearheading and coordinating research and technology efforts across the division, and integrating technology work more broadly across JPL. He has led several NASA, JPL and DARPA research and development activities, focused in the areas of spacecraft autonomy, digital engineering, and model-based software systems engineering. He received his Doctoral and Master's degrees from MIT's Department of Aeronautics and Astronautics, and his Bachelor's degree in Honours Mechanical Engineering from McGill University in Montreal, Canada.
\end{biographywithpic}

\begin{biographywithpic}
{Julie Castillo-Rogez}{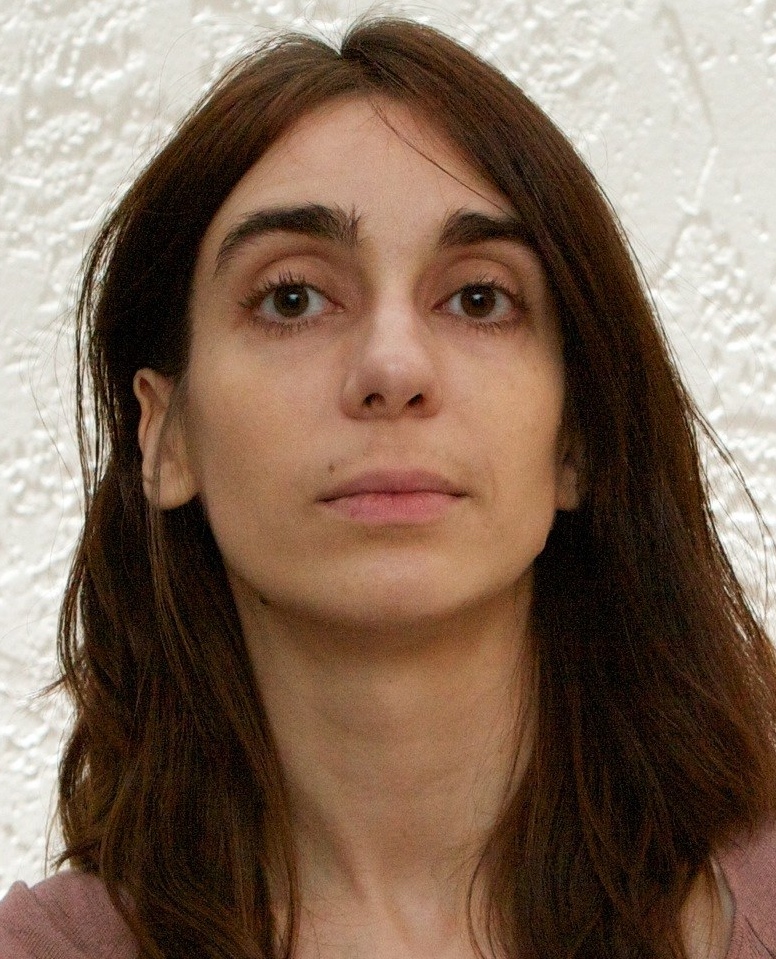}
received her B.S. degree in Geology in 1997 and a Ph.D. in Planetary Geophysics from the University of Rennes, France, in 2001. She is currently a planetary scientist at the Jet Propulsion Laboratory, California Institute of Technology. She has been the Project Scientist for the Dawn mission and is the Science Principal Investigator of the Near Earth Asteroid Scout mission. Her research interests include small-body geophysical and geochemical modeling and science autonomy. 
\end{biographywithpic}

\end{document}